# A Markov Chain Model for the Analysis of Round-Robin Scheduling Scheme

**D. Shukla**
Dept. of Mathematics and Statistics, Dr. H.S. Gour University, Sagar (M.P.), 470003, INDIA
Email:diwakarshukla@rediffmail.com
**Saurabh Jain**
Dept. of Comp. and Applications, Dr. H.S. Gour University, Sagar (M.P.), 470003, INDIA
Email:iamsaurabh_4@yahoo.co.in
**Rahul Singhai**
Dept. of Comp. and Applications, Dr. H.S. Gour University, Sagar (M.P.), 470003, INDIA
Email:singhai_rahul@hotmail.com
**Dr. R.K. Agarwal**
Indira Gandhi Engineering College, Sagar

-------------------------------------------------ABSTRACT-----------------------------------------------
In the literature of Round-Robin scheduling scheme, each job is processed, one after the another after giving a fix quantum. In case of First-come first-served, each process is executed, if the previously arrived processed is completed. Both these scheduling schemes are used in this paper as its special cases. A Markov chain model is used to compare several scheduling schemes of the class. An index measure is defined to compare the model based efficiency of different scheduling schemes. One scheduling scheme which is the mixture of FIFO and round robin is found efficient in terms of model based study. The system simulation procedure is used to derive the conclusion of the content

## 1. INTRODUCTION

In an operating system, a large number of processes arrive to the scheduler whose role is to manage the processing of these jobs. There are many scheduling schemes available in literature [see Silberschatz and Galvin [3], Stalling [7], Tanenbaum and Woodhull [8]] like FIFO, Round robin, Priority based, Multi-level queue and so on. All these have some advantages and disadvantages over each other. A unified study for scheduling scheme is required under a common environment. This motivates to design a general class of scheduling schemes so that its member may possess common properties of the class as well as could be mutually compared. With this thought of motivation, a general class of scheduling scheme is designed in this paper containing some well-known schemes like FIFO and Round robin as member schemes.

Shukla and Jain [4] have studied the multi-level queue-scheduling scheme in the environment of Markov chain model. Shukla et.al. [5] studied the setup of space division switches in a Markov chain model scenario. Shukla and Jain [6] used Markov chain model for deadlock-based study of multi-level queue scheduling. Some other related contributions are due to Medhi [1] and Naldi [2]. In the present study, the designed general class of scheduling scheme is examined through a Markov chain model in order to perform the comparative analysis of the performance of member scheduling schemes. The overall recommendation is that, under the Markov chain model based study, the general class contains scheme-III as the most recommendable.

## 2. GENERAL CLASS OF ROUND-ROBIN QUEUE SCHEDULING SCHEME

Consider a round-robin scheduling scheme shown in fig 2.1. A general class is laid down below:
(1) The S denotes scheduler and there are m processes $P_1, P_2, P_3,.. P_m$ in queue;
(2) The S provides one quantum of time to each process and next quantum is decided by a random trial;
(3) The S starts from any process $P_i$ in queue and then moves to $P_j$ $(j \neq i = 1,2,3...m)$;
(4) The new process enters from the end i.e. $P_{m+1}$ is placed after $P_m$ and so on;
(5) Suppose S is at any process $P_i$ (i=1, 2, 3…m) at the end of a quantum, then in the next quantum

(a) S will be on $P_{i+1}$ with priority p,
(b) S will be on $P_i$ with priority s,
(c) S will be on $P_{i-1}$ with priority q,

(6) The S becomes idle when there is no process in the queue. However it is assumed that the scheduler S may be in deadlock in any quantum;

(7) From this deadlock level, the S could be back also to the queue in any other quantum for processing purpose;

(8) There is a long waiting queue of processes $P_1$', $P_2$'….. outside the processing unit and if one process is over inside, then a new process, waiting outside, enters inside so as to maintain m processes there.

## 2.1 Proposed Markov Chain Model

Let $\{X^{(n)}, n \geq 1\}$ denotes a Markov chain with the state space $P_1, P_2, P_3 \ldots P_m$, D where D is a deadlock state used to denote idle, blocking or any disturbance caused in the system, during job processing. The $X^{(n)}$ is the state of scheduler of the system at the end of $n^{th}$ quantum (n=1,2,3…). Assume that m processes are in system at a time. Further let the transition of scheduler S is random over m+1 states in $n^{th}$ quantum. The transition diagram for any three processes $P_{i-1}, P_i, P_{i+1}$ and D is given in fig. 2.1. Define unit-step transition

$P[X^{(n+1)}=P_{i+1}/X^{(n)}=P_i] = p$
$P[X^{(n+1)}=P_i/X^{(n)}=P_i] = s$
$P[X^{(n+1)}=P_{i-1}/X^{(n)}=P_i] = q$
$P[X^{(n+1)}=D/X^{(n)}=P_i] = r$
$P[X^{(n+1)}=P_i/X^{(n)}=D] = 0$

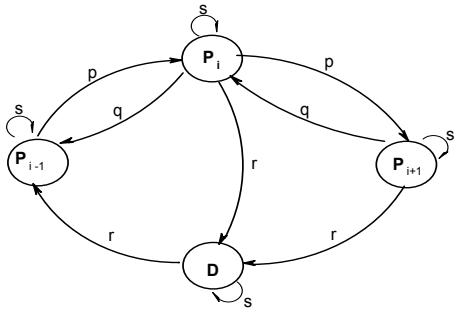

**Fig 3.1 (System Diagram)**

**Remark 2.1** The generalized expressions for n quantum are:
$P[X^{(n)}=P_i] = P[X^{(n-1)}=P_{i-1}].p + P[X^{(n-1)}=P_i].s +$
$\quad P[X^{(n-1)}=P_{i+1}].q$
$P[X^{(n)}=D] = \sum_{i=1}^{n} P[X^{(n-1)} = P_i]r + P[X^{(n-1)} = D]1$

## 3 Some Special Scheduling Schemes

By imposing restrictions and conditions over the ways and procedures, one can generate various scheduling schemes from the generalized class in section 2.1.

**3.1. Scheme- I [A]: When q = 0, p = 0, r=0, s = 1**

Then this general class has scheduling scheme FIFO for all quantum n.

**Remark 3.1.1.** The initial probabilities at n=0 for scheme-I[A] are:

$P[X^{(0)}=P_i]=pb_i$ and subject to the condition $\sum_{i=1}^{m} pb_i = 1$

**Remark 3.1.2.** The state probabilities after the first quantum are:
$P[X^{(1)}=P_i]=pb_i$

**Remark 3.1.3.** The generalized expressions of scheme-I [A] for n quantum are:
$P[X^{(n)}=P_i]=pb_i$

**3.2 Scheme- I [B]: When q = 0, p = 0, r+s = 1**

Then this general class has scheduling scheme FIFO for all quantum n.

**Remark 3.2.1** The initial probabilities at n=0 for scheme-I [B] are:
$P[X^{(0)}=P_i]=pb_i \qquad P[X^{(0)}=D]=0$

and subject to the condition $\sum_{i=1}^{m} pb_i = 1$

**Remark 3.2.2** The state probabilities after the first quantum are:

$P[X^{(1)}=P_i]=pb_i \cdot s \quad P[X^{(1)}=D]= r \cdot \sum_{i=1}^{m} pb_i = r$

**Remark 3.2.3** The state probabilities after the second quantum are:
$P[X^{(2)}=P_i]=P[X^{(1)}=P_i].s$
$P[X^{(2)}=D]= \sum_{i=1}^{m} P[X^{(1)} = P_i]r + P[X^{(1)} = D]1$

**Remark 3.2.4.** The generalized expressions of scheme-I [B] for n quantum are:

$P[X^{(n)}=P_i]=P[X^{(n-1)}=P_i].s$
$P[X^{(n)}=D]= \sum_{i=1}^{m} P[X^{(n-1)} = P_i]r + P[X^{n-(1)} = D]1$

**3.3 Scheme-II [A]: when q = 0, s = 0, r=0, p= 1**

This general class has scheme called "Round-Robin Scheduling scheme" for all quantum n.

**Remark 3.3.1.** The initial probabilities at n=0 for scheme-II [A] are:

$P[X^{(0)}=P_i]=pb_i$ and subject to the condition $\sum_{i=1}^{m} pb_i = 1$

**Remark 3.3.2.** The state probabilities after the first quantum are:
$P[X^{(1)}=P_i]=pb_{i-1}$

**Remark 3.3.3** The state probabilities after the second quantum are:
$P[X^{(2)}=P_i]=pb_{i-2}$

**Remark 3.3.4.** The generalized expressions of scheme-II [A] for n quantum are:
$P[X^{(n)}=P_i]=pb_{i-n}$

**3.4 Scheme- II [B]: When q = 0, s = 0, p + r = 1**

Then this general class has scheduling scheme called "Round-Robin Scheduling scheme" for all quantum n.

**Remark 3.4.1.** The initial probabilities at n=0 for scheme-I are
$P[X^{(0)}=P_i]=pb_i$ (i=1,2,3…m)
$P[X^{(0)}=D]=0$     and subject to the condition $\sum_{i=1}^{m} pb_i = 1$

**Remark 3.4.2.** The state probabilities after the first quantum are:
$P[X^{(1)}=P_i]=pb_{i-1}.p$     $P[X^{(1)}=D]=r.\sum_{i=1}^{m} pb_i = r$

**Remark 3.4.3.** The state probabilities after the second quantum are:
$P[X^{(2)}=P_i]=P[X^{(1)}=P_{i-1}].p$
$P[X^{(2)}=D]=\sum_{i=1}^{m} P[X^{(1)}=P_i].p.r + P[X^{(1)}=D].1$

**Remark 3.4.4.** The generalized expressions of scheme-II [B] for n quantum are:
$P[X^{(n)}=P_i]=P[X^{(n-1)}=P_{i-1}].p$
$P[X^{(n)}=D]=\sum_{i=1}^{m} P[X^{(n-1)}=P_i].p.r + P[X^{n-(1)}=D].1$

### 3.5 Scheme- III [A]: When q = 0, r=0, p + s= 1

**Remark 3.5.1.** The initial probabilities at n=0 for scheme-III [A] are:
$P[X^{(0)}=P_i]=pb_i$ and subject to the condition $\sum_{i=1}^{m} pb_i = 1$

**Remark 3.5.2.** The state probabilities after the first quantum are:
$P[X^{(1)}=P_i]=pb_{i-1}.p + pb_i.s$

**Remark 3.5.3.** The state probabilities after the second quantum are:
$P[X^{(2)}=P_i]=P[X^{(1)}=P_{i-1}].p + P[X^{(1)}=P_i].s$

**Remark 3.5.4.** The generalized expressions of scheme-III [A] for n quantum are:
$P[X^{(n)}=P_i]= P[X^{(n-1)}=P_{i-1}].p + P[X^{(n-1)}=P_i].s$

### 3.6 Scheme- III [B]: When q = 0, p + r + s = 1

**Remark 3.6.1** The initial probabilities at n=0 for scheme-III [B] are
$P[X^{(0)}=P_i]=pb_i$ (i=1,2,3…m)
$P[X^{(0)}=R]=0$ and subject to the condition $\sum_{i=1}^{m} pb_i = 1$

**Remark 3.6.2** The state probabilities after the first quantum are:
$P[X^{(1)}=P_i]= P[X^{(0)}=P_{i-1}].p + P[X^{(0)}=P_i].s$
$P[X^{(1)}=R]= r.\sum_{i=1}^{m} pb_i = r$

**Remark 3.6.3** The state probabilities after the second quantum are:
$P[X^{(2)}=P_i]=P[X^{(1)}=P_{i-1}].p + P[X^{(1)}=P_i].s$
$P[X^{(2)}=D]= \sum_{i=1}^{m} P[X^{(1)}=P_i]$

**Remark 3.6.4** The generalized expressions of scheme-III [B] for n quantum are:
$P[X^{(n)}=P_i]=P[X^{(n-1)}=P_{i-1}].p + P[X^{(n-1)}=P_i].s$
$P[X^{(n)}=D]= \sum_{i=1}^{m} P[X^{(n-1)}=P_i]$

### 3.7 Scheme- IV When q = 0, s = 0, r=0, p= 1

The general class has scheme called "Round-Robin Scheduling scheme" with condition that scheduler starts processing with first process for all quantum n.

**Remark 3.7.1** The initial probabilities at n=0 for scheme-II [A] are:
$P[X^{(0)}=P_i]=1$ (when i=1) and subject to the condition
$\sum_{i=1}^{n} pb_i = 1$

**Remark 3.7.2** The state probabilities after the first quantum are:
$P[X^{(1)}=P_i]=1$ (when i=2)

**Remark 3.7.3** The state probabilities after the second quantum are:
$P[X^{(2)}=P_i]=1$ (when i=3)

**Remark 3.7.4** The generalized expressions of scheme-IV for n quantum are:
$P[X^{(n)}=P_i]=1$ (when i=1)

### 6. SIMULATION STUDY

In order to compare all the four scheduling schemes under a common setup of Markov chain model, the simulation study is performed whose graphical output is below.

**Under Scheme-I[A]:**

Consider initial probabilities $pb_1=0.27$, $pb_2=0.15$, $pb_3=0.17$, $pb_4=0.18$, $pb_5=0.23$. Here p=q=r=0, s=1 and all $p_i$ are number of jobs.

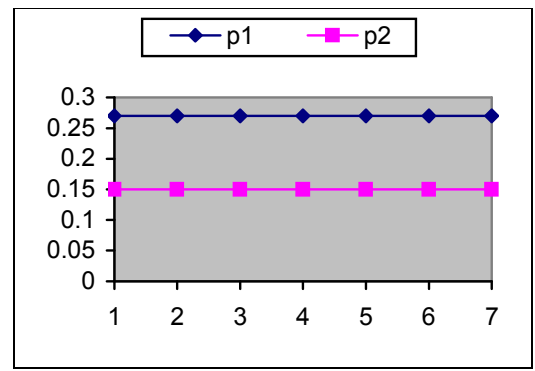

**Fig 4.1[A]**

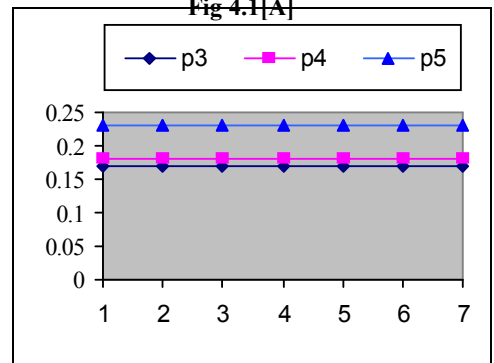

**Fig 4.1[B]**

In light of fig 4.1[A] and fig 4.1[B], this is to observe that the quantum variation does not affect the state probabilities $P_i$. The scheme-I[A] is purely first-come first-served (FIFO) with no chance of deadlock.

**Under Scheme-I[B]:**

Initial probabilities are $pb_1$ =0.27, $pb_2$ =0.15, $pb_3$=0.17, $pb_4$=0.18, $pb_5$=0.23, $pb_r$=0 with p = q = 0, r + s = 1 and r = 0.166

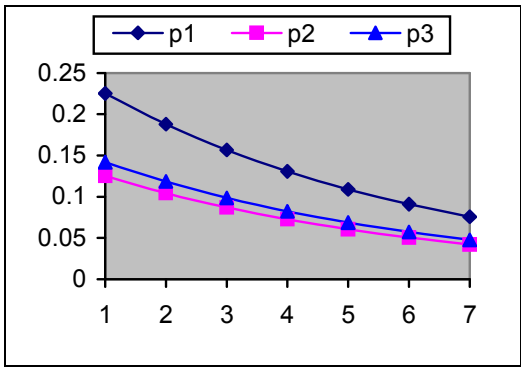

**Fig 4.2[A]**

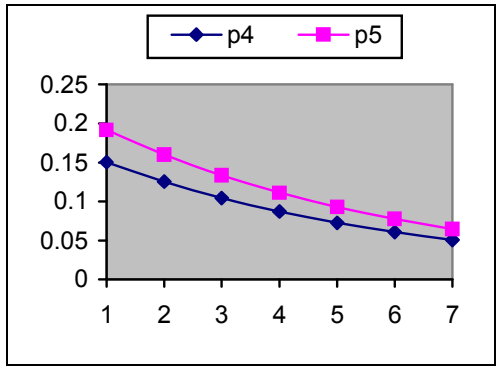

**Fig 4.2[B]**

Fig 4.2[A] and Fig. 4.2[B], relates to the same scheme but with the consideration of deadlock state. There is chance that during processing of jobs $P_i$ (i=1,2,3,4,5), the system may transit to state D and absorbed there. It found the probability that system survives on the same process over a large number of quantum reduces with a fast rate. This indicates for hanging chance of process scheduler if the process $P_i$ consumes more time. The chances of movement towards deadlock state are high for scheme I[B].

**Under Scheme-II[A]:**

Consider initial probabilities $pb_1$ =0.27, $pb_2$ =0.15, $pb_3$=0.17, $pb_4$=0.18, $pb_5$=0.23 with s=q=r=0, p=1.

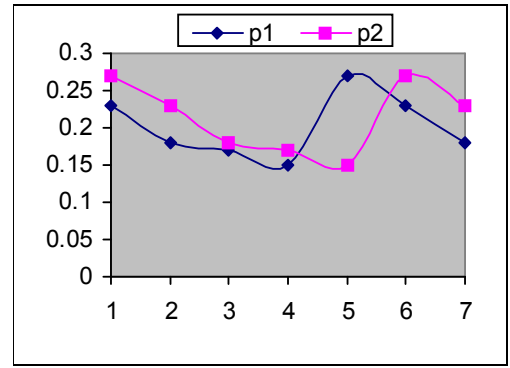

**Fig 4.3[A]**

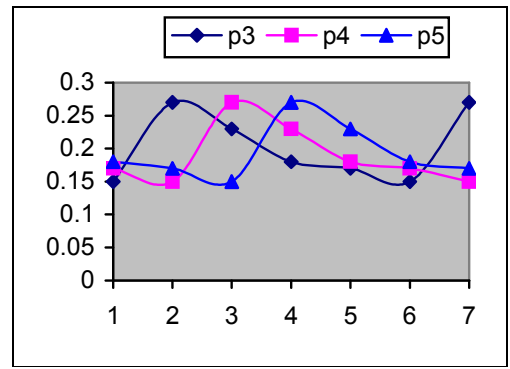

**Fig 4.3[B]**

In reference of fig. 4.3[A] and 4.3[B], the scheduling followed is round robin scheme with the condition that scheduler can start processing from any process with no deadlock condition. It is observed that at some specified quantum for an specified process, the probability attains a maximum. But over a large quantum, the downfall of probability occurs. At regular interval, after five quantum, the state probability bears a chance of scheduler being transited over the same.

**Under Scheme-II[B]:**

Initial probabilities are $pb_1$ =0.27, $pb_2$ =0.15, $pb_3$=0.17, $pb_4$=0.18, $pb_5$=0.23, $pb_r$=0 with s = q = 0, p + r = 1 and r = 0.166

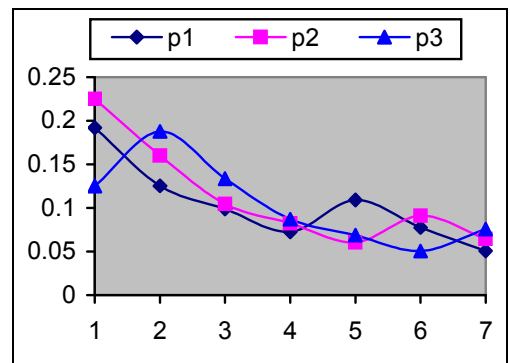

**Fig 4.4[A]**

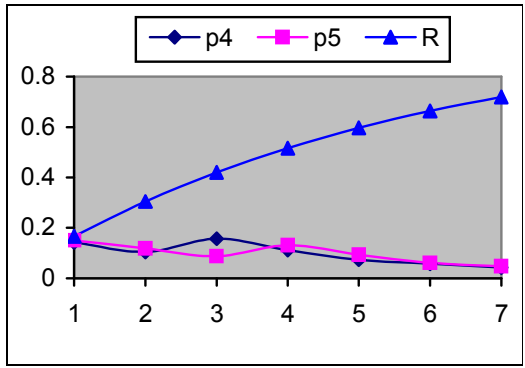

**Fig 4.4[B]**

When we refer to fig 4.4[A] and 4.4[B], where the scheme is round robin scheduling with the effect of deadlock state, the increasing number of quantum has indication for the system to be over the deadlock state.

**Under Scheme-III[A]:**

Consider initial probabilities $pb_1$ =0.27, $pb_2$ =0.15, $pb_3$=0.17, $pb_4$=0.18, $pb_5$=0.23 with s=0.5, p=0.5, q=r=0 and p + s = 1.

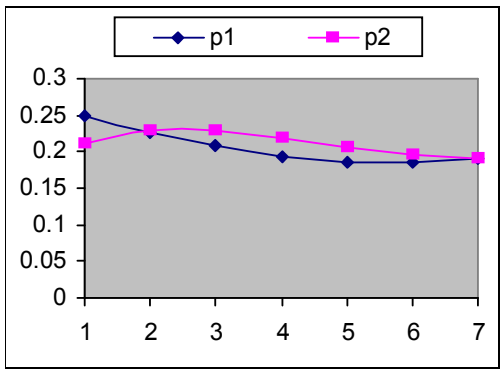

**Fig 4.5[A]**

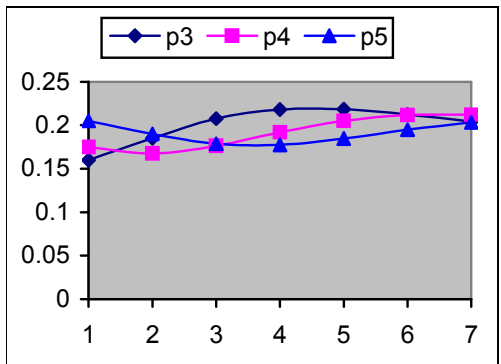

**Fig 4.5[B]**

The III[A] shown in fig 5.5[A] and 5.5[B] is neither FIFO nor a round robin scheme. But it is a mixture of these two. In this, the quantum distribution takes over the same state or to the next state depending upon the outcome of the random experiment. If the number of quantum increases, this scheme shows almost a stable pattern of the state probabilities. This means every process has almost same chance of being processed.

**Under Scheme-III[B]:**

Initial probabilities are $pb_1$ =0.27, $pb_2$ =0.15, $pb_3$=0.17, $pb_4$=0.18, $pb_5$=0.23, $pb_r$=0 with q = 0, p + r +s = 1 and r = 0.166

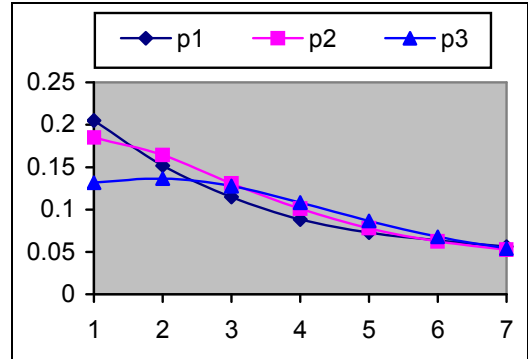

**Fig 4.6[A]**

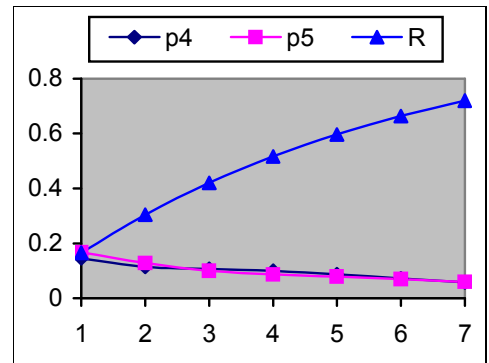

**Fig 5.6[B]**

**Under Scheme-IV:**

Consider initial probabilities $pb_1$ =1.0, $pb_2$ =0, $pb_3$=0, $pb_4$=0, $pb_5$=0 with p=1, q = r = s = 0.

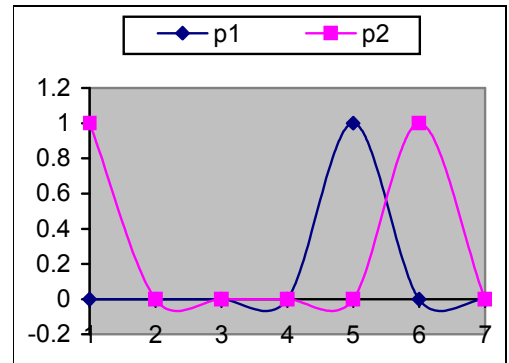

**Fig 4.7[A]**

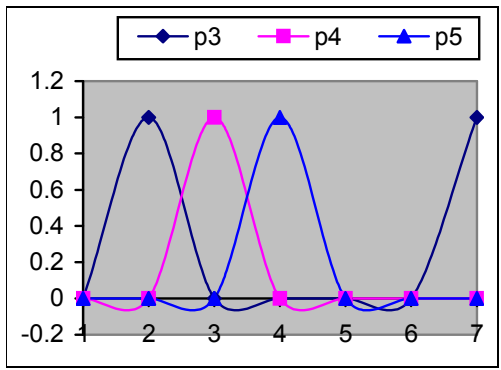

**Fig 4.7[B]**

The scheme-IV is purely a round robin scheme, which starts from the first process, the state probabilities are in fluctuating trend as evident from fig 4.7[A] and 4.7[B]. After a constant interval of quantum each process bears a high probability of being processed.

The overall view indicates that scheme-III bears more probability for processing jobs in comparison to other schemes. Because of its being a mixture of FIFO and round robin, the process scheduling aspect is stronger in this case.

6. **CONCLUDING REMARK**

The present study incorporates a general class of scheduling schemes with FIFO and round robin as its members. Some other schemes are also member of this class and all these are considered with and without deadlock state. All the schemes are examined through a common Markov chain model. The scheme-I is not as effective in comparison to others. In scheme-II[A], at the regular intervals after five quantum, the state probability bears a high chance of scheduler being transited over the same. If the number of quantum increases then scheme-III[A] shows almost a stable pattern of state probabilities. The scheme-III seems a good choice because of stability pattern over job processing under common Markov Chain Model setup.

**Author's Biography**

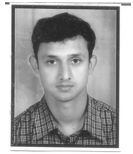 **Dr. Diwakar Shukla** is presently working as a faculty member in the Department of Mathematics and Statistics, H. S. Gour Sagar University, Sagar, M.P. and having over 19 years experience of teaching to U.G. and P.G. classes. He obtained M.Sc.(stat.), Ph.D.(stat.) degrees from Banaras Hindu University, Varanasi and served the Devi Ahilya University, Indore, M.P. as a permanent Lecturer from 1989 for nine years and obtained the degree of M.Tech.(Computer Science) from there. He joined Sagar University, Sagar as a Reader in statistics in the year 1998. During Ph.D. from BHU, he was junior and senior research fellow of CSIR, New Delhi through Fellowship Examination (NET) of 1983. Till now, he has published more than 55 research papers in national and international journals and participated in more than 35 seminars/conferences at the national level. He also worked as a Professor in the Lucknow University, Lucknow, U.P., for one (from june, 2007 to 2008) year and visited abroad to Sydney (Australia) and Shanghai (China) for conference participation and paper presentation. He has supervised seven Ph.D. theses in Statistics and Computer Science and eight students are presently enrolled for their doctoral degree under his supervision. He is a member of 10 learned bodies of Statistics and Computer Science at the national level. The area of research he works are Sampling Theory, Graph Theory, Stochastic Modeling, Data mining, Operation Research, Computer Network and Operating Systems.

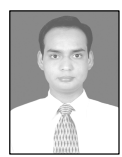 **Mr Saurabh Jain** has completed M.C.A. degree from Dr. H.S. Gour University, Sagar in 2005. He is presently working as Lecturer in the department of Comp. Science & Applications in the same University since 2007. He did his research in the field of Operating system. In this field, he has authored and co-authored 6 research papers in National/International proceedings and 3 research papers in National/International journals. His current research interest is to analyze the scheduler's performance under various algorithms.

**Mr. Rahul Singhai** has obtained M.C.A. degree from H.S. Gour University, Sagar, MP, in 2001 and obtained M.Phil degree in Computer Science from Madurai Kamaraj University, Madurai, Tamilnadu in 2008.. Presently he is pursuing Ph.D in Computer Science from H.S. Gour Central University, Sagar. His research interest includes Computer Network, Data mining

& Software Testing. He has authored and co-authored 8 research papers in proceedings & journals. Currently, he is working on to develop new imputation based methods for finding missing values in data preprocessing of data mining .He is appointed as a Lecturer in the Deptt. of Comp Science & Application, H.S. Gour Central University, sagar since 2005.

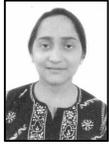

**Mrs. Shweta Ojha** received her M.C.A. degree from Dr. H.S. Gour University, Sagar in 2005. She is presently working as Lecturer in the department of Comp. Science & Applications in the same University since 2007. Her research interest is to analyze the scheduler's performance under various algorithms & Traffic analysis in various computer networking switches.. She has three research papers published in National/International conferences. Currently she is pursuing PhD in Computer Sc. From Dr. H.S.G. Central University, Sagar.